\documentstyle{article}
\newcommand{\beq}{\begin{equation}}
\newcommand{\eeq}{\end{equation}}
\newcommand{\beqa}{\begin{eqnarray}}
\newcommand{\eeqa}{\end{eqnarray}}
\newcommand{\n}{\nonumber\\}
\newcommand{\ddd}{\raisebox{2mm}{$\cdot$}\cdot\raisebox{-2mm}{$\cdot$}}
\def\Dot#1{\stackrel{\raisebox{-1mm}{\boldmath $\cdot$}}{#1}}
\begin{document}
\begin{titlepage}
\title{
\bf
The Painlev\'{e} Property, W Algebras \\and \\Toda Field Theories\\
associated with \\Hyperbolic Kac-Moody Algebras
}
\author{
Reinhold W. Gebert$^a$
\thanks{Supported by Deutsche Forschungsgemeinschaft under Contract
              Ni 290/3-1.},
Takeo Inami$^b$
\thanks{Supported partially by Grant-in-aid of Japanese
Ministry of Education, Science and Culture, Nos. 06221241 and 06044257.},
and
Shun'ya Mizoguchi$^a$
\thanks{Supported by the Alexander von Humboldt Foundation.}\\
\\
\and
$a$: II. Institut f\"{u}r Theoretische Physik,
            Universit\"{a}t Hamburg,\\
            Luruper Chaussee 149, 22761 Hamburg, Germany
\and
$b$: Yukawa Institute for Theoretical Physics, \\Kyoto University,
Kyoto 606-01, Japan
}
\date{March, 1995}
\maketitle
\begin{abstract}
We show that the Painlev\'{e} test is useful not only for probing
(non-)\\ integrability but also for finding the values of spins of
conserved currents (W currents) in Toda field theories (TFTs). In
the case of the TFTs based on simple Lie algebras
the locations of resonances are shown to give precisely
the spins of conserved W currents.
We apply this test to TFTs based on strictly hyperbolic Kac-Moody
algebras and show that there exist no resonances other than that at
$n=2$, which corresponds to the energy-momentum tensor, indicating
their non-integrability.
We also check by direct calculation that there are no spin-3 nor -4
conserved currents for all the hyperbolic TFTs
in agreement with the result of our Painlev\'{e} analysis.
\end{abstract}\end{titlepage}

\section{Introduction}
Toda field theories define integrable field theories if they
are associated with a Cartan matrix of a simple Lie algebra
or a affine Kac-Moody algebra
\cite{LS1,Flaschka1,LS2,Kostant,MOP,WilsonErg,Mansfield,OliveTurok}.
They have broad application in mathematical and theoretical physics,
and in particular have attracted particle physicists' interest
in the various areas of researches
\cite{Smatrix,HM,perturbation,deVega,Hollowood,OTU,recent}.

A Toda field theory is governed by the Lagrangian
\beq
{\cal L}=\frac{1}{2}\sum_{i,j=1}^N K_{ij}
\partial_+\varphi_i\partial_-\varphi_j
+\frac{1}{\beta}\sum_{i=1}^N\exp\left(
\beta\sum_{j=1}^N K_{ij}\varphi_j
\right),
\eeq
where $K_{ij}$ is a Cartan matrix of some Lie algebra \mbox{\boldmath $g$}.
It is referred to as conformal and affine Toda
field theory, respectively, depending on whether \mbox{\boldmath $g$}
is a simple Lie algebra or a affine Kac-Moody algebra.
The former is a generalization of the Liouville theory. It
is conformally invariant, admitting no soliton solutions.
The latter is a relative of the sine-Gordon theory.
The fields then become massive and the conformal invariance is lost.
The addition of the extra potential term associated with the
highest root can be regarded as a deformation of a conformal
field theory \cite{dCFT}. The field equations have soliton solutions
if the coupling constant is imaginary \cite{Hollowood}.

A more general class of Kac-Moody algebras can be defined
in association with {\it generalized Cartan matrices} \cite{Kac},
which do not necessarily provide positive (semi-)definite root spaces.
It is then interesting to ask what property the Toda theory will
possess, if \mbox{\boldmath $g$} is taken to be one in this class
of Kac-Moody algebra. In this paper we will consider the case of
{\it hyperbolic Kac-Moody algebra} ({\it hyperbolic Toda field
theory, HTFT}), mainly focusing on the issue of its
(non-)integrability.

One of the reasons why we are interested in the integrability
of HTFTs concerns
the existence of W currents \cite{Walgebra}. Integrability of a
Toda field theory reflects the existence of as many conserved
currents as the number of degrees of freedom in general \cite{FL}.
It was shown that one can reconstruct exact solutions of Toda
field equations from solutions of Miura-type differential equations
\cite{BG1}. Since the conserved currents of spin-2 and higher associated
with conformal Toda field theories are known to generate a W algebra
\cite{BG2,O'Raifeartaigh}, one may expect that, if HTFTs
are integrable, one may then obtain a new class of W algebras
through their conserved currents which are supposed to exist.
This will open up a new direction to extend a symmetry of
conformal field theory and string theory.
We hope our study will shed light from a physical point
of view on the nature of hyperbolic Kac-Moody algebras, which has
been scarcely understood as yet.

It is not easy in general to prove whether or not a given field
theory is integrable (or non-integrable) in a rigorous sense.
A practical method has been contrived for this aim by Weiss
et.~al.~\cite{Weiss1,Weiss2}, in which so-called `the Painlev\'{e}
property' was utilized to probe integrability of partial
differential equations (See \cite{RGB} for a review.).
It was demonstrated there that such
analyses work very well in a number of integrable models.
We call this `experimental' test {\em the Painlev\'{e} test}.
Applying this test to HTFTs, we show that the strictly hyperbolic
Toda field theories (SHTFT) do not pass the test in the following
sense: The minimal number of arbitrary functions that a generic
solution of the Toda field equation possesses is smaller than that
of ordinary integrable conformal Toda field theories if the
solution is expanded around a singular manifold.

In the subsequent sections we will first `rediscover' that,
in the case of simple Lie algebras, resonances occur
precisely at $n=\mbox{exponents}+1$. This relation was
reported earlier by Flaschka and Zeng \cite{FlaschkaZeng}.
We will give an alternative, direct proof to the key theorem for
the relation. We will next show that a unique resonance occurs at
$n=2$ in the case of SHTFTs, suggesting that only a single conserved
current (the energy-momentum tensor) exists for these theories.
We will also check explicitly that there are indeed no spin-3 and -4
conserved currents for any HTFT, which supports the result
of our Painlev\'{e} analysis.

The plan of this paper is as follows. In sect.2 we will give a brief
review of the Painlev\'{e} test. We will devote sect.3 to the
application of the test to the HTFTs. After the definition of the
hyperbolic Kac-Moody algebra, we will prove that any Toda field
theory associated with an invertible Cartan matrix is conformally
invariant \cite{Sasakietal}. We will then see a beautiful relation between
the resonances and the exponents of simple Lie algebras, and show further
that any SHTFT does not pass the Painlev\'{e} test. In sect.4 we will
check the non-existence of conserved currents of spin-3 and -4 for
any HTFT. Finally we will summarize our results and future prospects
in sect.5. Appendix A and B contain proofs of the Remark.2 and
the Theorem, respectively.

\section{Review of the Painlev\'{e} test}

A system of ordinary differential equations are said to possess
the Painlev\'{e} property if all its `movable' singularities
(singularities whose locations depend on the initial conditions)
are pole singularity. The first observation on the relation between
the Painlev\'{e} property and integrability was made by
S.Kowalevskaya in 1889
in her work of rigid-body problems \cite{Kowalevskaya}.
This property is named after
Painlev\'{e}, who classified the second-order differential equations
which possess such a property.

Weiss et.~al.~generalized the notion of the Painlev\'{e} property
to partial differential equations of $N$ complex variables
($z_1,\ldots, z_N$). They assumed the form of the solutions as
\beq
u(z_1,\ldots,z_N)=\phi^{\alpha}\sum_{n=0}^{\infty} u_n\phi^n
\label{uexpansion}
\eeq
in a neighborhood of a `singular manifold'
\beq
\phi=\phi(z_1,\ldots,z_N)=0,
\eeq
and said that the partial differential equation has the Painlev\'{e}
property when the expansion coefficients $u_n$ consistently
contain $2N-1$ arbitrary functions. They showed that a wide variety
of known integrable equations,
e.g. KdV, MKdV,
Boussinesq, higher-order KdV and KP equations, enjoy this property
\cite{Weiss1,Weiss2}.
\vskip 5mm\noindent
{\it Example.}~~~
Burger's equation
\beq
\Dot{u} +uu'=\sigma u''
\label{Burger}
\eeq
($\Dot{\phantom{u}}$ and $\phantom{u}'$ denote
$\frac{\partial}{\partial t}$  and
$\frac{\partial}{\partial x}$, respectively.)
is, substituting the expansion (\ref{uexpansion}) into
(\ref{Burger}), reduced to the recursion relation of $u_n$:
\beqa
&&\Dot{u}_{n-1}+(n-2)u_{n-1}\Dot{\phi}
+\sum_{m=0}^n u_{n-m}[u_{m-1}'+(m-1)\phi'u_m]\n
&=&\sigma[u_{n-2}''+2(n-2)u_{n-1}'\phi'
+(n-2)u_{n-1}\phi'' +(n-1)(n-2)u_n(\phi')^2].
\label{Burgerreceq}
\eeqa
Here $\alpha$ is determined to be $-1$ by the leading-order analysis.
It turns out that (\ref{Burgerreceq}) holds
identically at $n=2$, and hence $u_2$ cannot be determined but
is regarded as an arbitrary function. The values of $n$ at which
the recursion relation is identically satisfied (and hence there is
a room for an arbitrary function) as above are called
`resonances' \cite{Weiss0}.
Integrability then requires $2N-1$ resonances.
\vskip 5mm

It was shown by Yoshida
\cite{Yoshida} that the integrable Toda field theories
(`generalized Toda lattice')
based on simple Lie algebras are strongly characterized by their
Painlev\'{e} property (See also \cite{AM}).
Flaschka and Zeng were the first to show
the correspondence between the locations of resonances and the exponents
of the Lie algebra \cite{FlaschkaZeng}. This relation was also known
to Yoshida \cite{Yoshidaproblem}. We found this relation independently;
we will give another elementary proof of this, and generalize the
analysis to HTFTs.

\section{Painlev\'{e} test for SHTFTs}
\subsection{Hyperbolic Kac-Moody algebras}

We now give a brief description of hyperbolic Kac-Moody algebras
\cite{Kac}.
An $N\times N$ matrix $K_{ij}$ is a generalized Cartan matrix
if it satisfies (i) $K_{ij}\in\mbox{\bf Z}$, (ii) $K_{ii}=2$,
(iii) $K_{ij}\leq 0$ $(i\neq j)$ and (iv) $K_{ij}=0$ if $K_{ji}=0$.
One can define a generalized Kac-Moody algebra by the generating
relations
\begin{equation}
{[}h_i, h_j{]}=0,~
{[}h_i, e_j{]}=K_{ij}e_j,~
{[}h_i, f_j{]}=-K_{ij}f_j,~
{[}e_i, f_j{]}=\delta_{ij}h_j
\end{equation}
together with the Serre relations
\begin{equation}
({\rm ad}e_i)^{1-K_{ij}}(e_j)=0,~
({\rm ad}f_i)^{1-K_{ij}}(f_j)=0
\end{equation}
for $i\neq j$. A generalized Kac-Moody algebra is said hyperbolic
Kac-Moody algebra (strictly hyperbolic Kac-Moody algebra, respectively)
if the associated Dynkin diagram is of hyperbolic type (strictly
hyperbolic type), i.e. if any deletion of nodes from the diagram
leaves a direct sum of those of finite or affine type (finite type
only). We will also use the same terminology for a Cartan matrix $K$.

Hyperbolic Kac-Moody algebras for $7\leq\mbox{rank}\leq 10$ were
first classified in \cite{Kac}. The list of all the 136 hyperbolic
Kac-Moody algebras for $3\leq\mbox{rank}\leq 10$ was given in
\cite{Sac}. Together with all rank-2 generalized Kac-Moody algebras
associated with Cartan matrices in the form \cite{LM}
\beq
K=\left[
\begin{array}{cc}
2&-k\\
-l&2
\end{array}
\right],~~~k,l\in {\mbox{\bf Z}},~~~  kl > 4,~~~ k,l > 0,\label{rank2}
\eeq
they exhaust all hyperbolic Kac-Moody algebras.
These Kac-Moody algebras have root
spaces with Lorentzian signature, and hence a HTFT
contains a single ghost-like field.

It is only very recently that a few mathematicians have begun
representation theoretic studies of generalized Kac-Moody algebras
\cite{B,FFR} and that physicists have looked for applications of
these algebras in particle physics (e.g. \cite{Nicolai}).
The most familiar example of hyperbolic Kac-Moody algebras
will be $E_{10}$, which has the maximal allowed rank,
in the context of string compactification
(See \cite{GN,G} for recent aspects on this subject.).

\subsection{Conformal invariance of non-affine Toda field theories
\mbox{\rm [27]}
}

In the light-cone coordinates the Toda equation of motion is given by
\beq
\partial_+\partial_-\varphi_i=\exp\left(
\beta\sum_{j=1}^N K_{ij}\varphi_j
\right).\label{EofM}
\eeq
Under a conformal transformation
\beq
x^{\pm}\rightarrow
\overline{x}^{\pm}(x^{\pm})
\eeq
the LHS of (\ref{EofM}) changes as
\beq
\partial_+\partial_-\varphi_i
\rightarrow
\overline{\partial}_+\overline{\partial}_-\varphi_i
=
\frac{\partial x^+}{\partial \overline{x}^+}
\frac{\partial x^-}{\partial \overline{x}^-}
\partial_+\partial_-\varphi_i.
\eeq
The invariance of the RHS of (\ref{EofM}) then requires that
\beq
\varphi_i\rightarrow\overline{\varphi}_i
=\varphi_i+\frac{\lambda_i}{\beta}\ln\left(
\frac{\partial x^+}{\partial \overline{x}^+}
\frac{\partial x^-}{\partial \overline{x}^-}
\right)
\eeq
for some $\lambda_i$ such that
\beq
\sum_{j=1}^N K_{ij}\lambda_j=1
\eeq
for any $i$. The solution is
\beq
\lambda_i=\left(
K^{-1}\left[
\begin{array}{c}1\\
\vdots\\
1
\end{array}
\right]
\right)_i.
\eeq
Hence one may find $\lambda_i$ for any TFTs except affine TFTs.
In other words, all but affine TFTs are conformally invariant. This
fact implies the existence of a conserved chiral energy-momentum
tensor in every non-affine TFTs.

In the case of the TFTs corresponding to simple Lie algebras,
$\lambda_i$ is related to `half the sum of positive roots'
(the Weyl vector). The above may be considered as a generalization
(or a `regularization') of this notion for general TFTs.

\subsection{The Painlev\'{e} test}

We now apply the Painlev\'{e} test to Toda field theories. We will
closely follow ref.\cite{Yoshida} (See Remark.1 below, however.).
The Toda field equations are given by
\beq
\frac{\partial^2 \varphi_i}{\partial x\partial t}
=-\exp\left(\sum_{j=1}^N K_{ij}\varphi_j
\right)~~~(i=1,\ldots,N).
\label{TFeq}
\eeq
We have omitted the coupling constant since it has no relevance here.
Eq.(\ref{TFeq}) can be cast into an equivalent system of first order
differential equations
\beqa
\frac{\partial}{\partial t}A_i&=&A_i\sum_{j=1}^N K_{ij}B_j
{}~~~(i=1,\ldots,N),\n
\frac{\partial}{\partial x}B_j&=&-A_j=-\sum_{i=1}^N \delta_{ij}A_i
{}~~~(i=1,\ldots,N).
\label{ls}
\eeqa
\vskip 5mm\noindent
{\it Remark.1.}~~~
Note that in ref.\cite{Yoshida} a slightly different
system of equations
\beq
\Dot{A}^{\raisebox{-2.5mm}{\scriptsize$j$}}
=A^j\sum_{i=1}^N(D\eta)_j^{~i}B_i,~~~
B_i'=-\sum_{j=1}^ND_{ji}A^j
\eeq
is adopted,
where $\eta^{ij}=\delta^{ij}$ and $D_{ij}$ is a matrix whose rows
consist of simple roots (different from eq.(\ref{D})). Our choice
(\ref{ls}) has an advantage in the hyperbolic case in that the
information on the signature of the root space is encoded only in
the Cartan matrix $K_{ij}$, and hence we need not care about
raising and lowering the indices by an indefinite metric.
\vskip 5mm

Following the usual prescription of the Painlev\'{e} test,
we assume that the solutions of
(\ref{ls}) are single-valued around some singular manifold
$\phi(x,t)=0$. Substituting the expansion
\beq
A_i=\phi^{-n_A}\sum_{n=0}^{\infty}A_i^{(n)}\phi^n,~~~
B_j=\phi^{-n_B}\sum_{n=0}^{\infty}B_j^{(n)}\phi^n
\label{ABexpansion}
\eeq
into (\ref{ls}), we have a recursion relation
\beq
T^{(n)}\vec{X}^{(n)}=\vec{b}^{(n)}\label{rec}
\eeq
for the expansion coefficients
$
\vec{X}^{n}\equiv(A_1^{(n)},\ldots,A_N^{(n)},
B_1^{(n)},\ldots,B_N^{(n)})^T
$
($T$ denotes the transpose.), where $T^{(n)}$ and $\vec{b}^{(n)}$
are given by
\beq
T^{(n)}=\left[
\begin{array}{cc}P^{(n)}&Q^{(n)}\\
R^{(n)}&S^{(n)}\end{array}
\right],~~~~~~
\vec{b}^{(n)}=\left[
\begin{array}{c}b_i^{(n)}\\
b_{N+j}^{(n)}\end{array}
\right],
\eeq
\beqa
P_{ik}^{(n)}&=&
\left(
(n-2)\Dot{\phi}-\sum_{j=1}^N K_{ij}B_j^{(0)}
\right)\delta_{ik},\n
Q_{il}^{(n)}&=&
-A_i^{(0)}K_{il},\n
R_{jk}^{(n)}&=&
\delta_{jk},\n
S_{jl}^{(n)}&=&
(n-1)\phi'\delta_{jl},\n
b_i^{(n)}&=&-\Dot{A}_i^{\raisebox{-1.5mm}{\scriptsize $(n-1)$}}
+\sum_{m=1}^{n-1}A_i^{(n-m)}\sum_{j=1}^N K_{ij}B_j^{(m)},\n
b_{N+j}^{(n)}&=&-{B_j^{(n-1)}}'
\eeqa
$(i,j,k,l=1,\ldots,N)$.

\vskip 5mm\noindent
{\it Remark.2.}~~~
$n_A$ and $n_B$ can be shown to satisfy
\beq
n_A=n_B+1.
\eeq
The Laurent series (\ref{ABexpansion}) is called `balance', and
in particular is called `lowest balance' if \cite{Flaschka2}
\beq
A_i^{(0)}\neq 0~~~\mbox{for any $i=1,\ldots,N$}.
\label{allAnonzero}
\eeq
It can be shown that for the TFTs based on either simple
Lie algebras or strictly hyperbolic Kac-Moody algebras
the only possibility is
\beq
n_A=2,~~~n_B=1, \label{na=nb+1=2}
\eeq
while for general generalized Kac-Moody algebras
\beq
n_A=n_B+1\geq 3
\eeq
are allowed in general. However, (\ref{na=nb+1=2}) is only possible
for lowest balances in the latter case as well, as far as $K$ is
invertible. The proof is given in Appendix A. In this paper we restrict
ourselves to the lowest balances (and hence the case (\ref{na=nb+1=2})
only).
\vskip 5mm

Evidently $\mbox{det}T^{(n)}$ must vanish if $n$ is a resonance,
so let us calculate the determinant of $T^{(n)}$.
Due to the assumption eq.(\ref{rec}) for $n=0$ reads
\beq
B_i^{(0)}=-2\Dot{\phi}\sum_{j=1}^N(K^{-1})_{ij},~~~
A_i^{(0)}=-2\Dot{\phi}\phi'\sum_{j=1}^N(K^{-1})_{ij}
\eeq
$(i=1,\ldots,N)$. We write $T^{(n)}$ explicitly as
\beq
T^{(n)}=\left[
\begin{array}{ccccccc}
n\Dot{\phi}&&          & &-A_1^{(0)}K_{11}&\cdots&-A_1^{(0)}K_{1N}\\
           &\ddd&\mbox{\LARGE $0$}& &        \vdots&\cdots&\vdots\\
           &\mbox{\LARGE $0$}&\ddd& &        \vdots&\cdots&\vdots\\
           &&&n\Dot{\phi}&-A_N^{(0)}K_{N1}&\cdots&-A_N^{(0)}K_{NN}\\
1          &&          & &(n-1)\phi'&   &\\
          &\ddd&\mbox{\LARGE $0$}& &&\ddd~~~\mbox{\LARGE $0$}   &\\
          &\mbox{\LARGE $0$}&\ddd& &&\mbox{\LARGE $0$}~~~\ddd   &\\
          &&&1 &&   &(n-1)\phi'
\end{array}
\right]~.
\eeq
It is easy to see that the determinant is given by
\beq
\mbox{det}T^{(n)}=(\Dot{\phi}\phi')^N
\det[
n(n-1)\cdot{\bf 1}-2DK
],
\eeq
where
\beq
D\equiv\left[
\begin{array}{ccc}
\sum_{j=1}^N(K^{-1})_{1j}&&\\
&\ddd&\\
&&\sum_{j=1}^N(K^{-1})_{Nj}
\end{array}
\right].
\label{D}
\eeq
Hence $\mbox{det}T^{(n)}=0$ is equivalent to the characteristic equation
for the matrix $2DK$ of eigenvalues $n(n-1)$. In other words, the
calculation of resonances is deduced to an eigenvalue problem.

\newcommand{\ld}{\lambda}
\newcommand{\lb}{(}
\newcommand{\rb}{)}
We first give the results for the simple Lie algebras.
\vskip 5mm\noindent
{\bf Proposition.1.}~~~
{\it
Set $\ld=n(n-1)$, then $\mbox{det}T^{(n)}$ is given by
$(\Dot{\phi}\phi ')^N\times$
\beqa
A_{N}&:&\lb\ld -1\cdot 2\rb\lb\ld -2\cdot 3\rb\lb\ld -3\cdot 4\rb\cdots
\lb\ld -N(N+1)\rb,\n
D_{2M}&:&\lb\ld -1\cdot 2\rb\lb\ld -3\cdot 4\rb\lb\ld -5\cdot 6\rb
\cdots \lb\ld -(2M-3)(2M-2)\rb\n
&&\cdot
\lb\ld -(2M-1)2M\rb^2\lb\ld -(2M+1)(2M+2)\rb
\n
&&\cdots
\lb\ld -(4M-3)(4M-2)\rb,\n
D_{2M+1}&:&\lb\ld -1\cdot 2\rb\lb\ld -3\cdot 4\rb\lb\ld -5\cdot 6\rb
\cdots \n
&&\hskip -1mm\cdot
\lb\ld -(2M-1)2M\rb\lb\ld -2M(2M+1)\rb\lb\ld -(2M+1)(2M+2)\rb\n
&&\cdots
\lb\ld -(4M-1)(4M)\rb,\n
B_N(C_N)&:&\lb\ld -1\cdot 2\rb\lb\ld -3\cdot 4\rb\lb\ld -5\cdot 6\rb\cdots
\lb\ld -(2N-1)2N\rb,\n
E_6&:&\lb\ld-2\rb\lb\ld- 20\rb\lb\ld- 30\rb
\lb\ld- 56\rb\lb\ld- 72\rb\lb\ld- 132\rb,\n
E_7&:&\lb\ld-2\rb\lb\ld- 30\rb\lb\ld- 56\rb
\lb\ld- 90\rb\lb\ld- 132\rb\lb\ld- 182\rb\lb\ld- 306\rb,\n
E_8&:&\lb\ld-2\rb\lb\ld- 56\rb\lb\ld- 132\rb\lb\ld- 182\rb
\lb\ld- 306\rb\lb\ld- 380\rb\lb\ld- 552\rb\n
&&\cdot\lb\ld- 870\rb,\n
F_4&:&\lb\ld-2\rb\lb\ld- 30\rb\lb\ld- 56\rb\lb\ld- 132\rb,\n
G_2&:&\lb\ld-2\rb\lb\ld- 30\rb.
\eeqa
}
\vskip 5mm

The number appearing in each factor is always a product of two consecutive
integers, the larger one of which corresponds to a (possible) resonance.
Remarkably, we find that the (possible) resonances occur precisely at
the values
\beq
n=\mbox{exponents}+1
\eeq
for any Toda field theory based on a
simple Lie algebra (Table 1). For simple Lie algebras with small rank we
have checked the compatibility as well. We also notice that the resonances
occur not $2N$ times, but only $N$ (=rank) times. This fact, which may be
seen as a discrepancy, has been known for some time, being interpreted
as a limit where the missing $N-1$ arbitrary functions go to infinity
\cite{Flaschka2}.

This relation between the locations of the resonances
and the exponents was first discovered by Flaschka and Zeng
\cite{FlaschkaZeng}. The Proposition.1
is a consequence of the following Theorem:
\vskip 5mm\noindent
{\bf Theorem.}~~~
{\it
Let $K$ be a Cartan matrix of a simple Lie algebra \mbox{\boldmath $g$}
of rank $N$, and
$D$ be a diagonal matrix defined in (\ref{D}).
Let $\{\chi_i|~i=1,\ldots,N \}$ be the set of
exponents of \mbox{\boldmath $g$}, then the set of eigenvalues of
the matrix $2KD$ is given by $\{\chi_i(\chi_i+1)|~i=1,\ldots,N \}$.}
\vskip 5mm
In ref.\cite{FlaschkaZeng} the above Theorem was proven by invoking
the property of the Casimir of the principal sl(2) of
\mbox{\boldmath $g$} \cite{Varadarajan}.
In Appendix B we will give an
alternative, direct proof of the Theorem by means of induction
for completeness.

Since the values of conserved W currents and
the exponents of the Lie algebra are in one-to-one correspondence
\cite{O'Raifeartaigh}, the above result implies that the Painlev\'{e}
test may tell us not only about its integrability, but also more
detailed information about the spins of existing conserved currents.
The Painlev\'{e} test can then be a powerful tool to search for new
conserved W currents for Toda field theories based on generalized
Kac-Moody algebras.

If this is true, then the Painlev\'{e} test should `detect' the conserved
energy-momentum tensor, which exists in any Toda field theory. In fact
this is the case.
\vskip 5mm\noindent
{\bf Proposition.2.}~~~
The matrix
$DK$ and hence $KD$
always has an eigenvalue 1.
\vskip 5mm\noindent
{\it Proof.}~~~
The latter is given by
\beq
KD=\left[
\begin{array}{ccc}
K_{11}\sum_{j=1}^N(K^{-1})_{1j}&\cdots&K_{1N}\sum_{j=1}^N(K^{-1})_{Nj}\\
&\cdots&\\
K_{N1}\sum_{j=1}^N(K^{-1})_{1j}&\cdots&K_{NN}\sum_{j=1}^N(K^{-1})_{Nj}
\end{array}
\right].
\eeq
Hence
\beqa
KD\left[
\begin{array}{c}
1\\ \vdots\\ 1\end{array}
\right]
&=&
\left[
\begin{array}{c}
\sum_{k=1}^NK_{1k}\sum_{j=1}^N(K^{-1})_{kj}\\
\vdots\\
\sum_{k=1}^NK_{Nk}\sum_{j=1}^N(K^{-1})_{kj}
\end{array}
\right]\n
&=&
\left[
\begin{array}{c}
\sum_{j=1}^N \delta_{1j}\\
\vdots\\
\sum_{j=1}^N \delta_{Nj}
\end{array}
\right]
=1\cdot
\left[
\begin{array}{c}
1\\ \vdots\\ 1\end{array}
\right].
\eeqa
This shows that $[1,\ldots,1]$ is always an eigenvector of $KD$ of
eigenvalue 1, no matter what the Cartan matrix is (as far as it is
invertible). q.e.d.
\vskip 5mm

We have thus seen that there is always a resonance at $n=2$
for any TFTs. We have also checked the compatibility of the
recursion relation for rank-2 SHTFTs discussed below.


We will now apply the test to the HTFTs.
In this paper we restrict our analysis to the SHTFTs
(See sect.5, however.).
The Cartan matrices of the strictly hyperbolic Kac-Moody
algebras are classified into two classes. The first class consists of
those for all rank-2 hyperbolic Kac-Moody algebras. They are
of the form (\ref{rank2}),
and are infinite in number. The second class consists of those
associated with the eleven Dynkin diagrams listed up in the Table 2.
In this case the rank is either three or four.

For the first class the determinant of $T^{(n)}$ is
calculated as
\beq
\mbox{det}T^{(n)}
=\Dot{\phi}\phi'(n+1)(n-2)\left[
n(n-1)-2\frac{(2+k)(2+l)}{4-kl}
\right]. \label{detrank2}
\eeq
Due to(\ref{rank2}), $k,l$ satisfy
\beq
\mbox{det}K=4-kl<0,~~~k,l>0, \label{kl>4}
\eeq
which means that
\beq
-2\frac{(2+k)(2+l)}{4-kl}>0.
\eeq
Hence the two solutions $n$ of the equation
$[\cdots]~\mbox{(in (\ref{detrank2}))}=0$ are either both positive
or both negative. At the same time the sum of them must be 1.
Clearly there exist no two integers that satisfy both
requirements. This shows that the only positive integer solution
of $(\ref{detrank2})=0$ is $n=2$.

For the second class, on the other hand, we have explicitly checked
for all the eleven cases that the matrix $2KD$ does not have any
positive integer eigenvalue other than 2. The result is summarized in
Table 3.

The Painlev\'{e} test
thus suggests that the SHTFTs have no conserved currents except the
energy-momentum tensor and that they are non-integrable.

\section{Search for conserved currents in HTFTs}

As we already discussed in Introduction, the conserved currents which
are as many as the number of degrees of freedom in a conformal TFT
are a consequence of their integrability (and vice versa). In this
section we will study HTFTs from this point of view.

\subsection{Spin 3}

The equations of motion are given by (\ref{EofM}).
Due to the conformal invariance,
we only  consider the chiral ($x^{+}$-dependent) currents.
The most general form of the spin-3 current is
\beq
W^{(3)}\equiv
\sum_{i,j,k=1}^{N}
a_{ijk}\partial_+\varphi_i\partial_+\varphi_j\partial_+\varphi_k
+\sum_{i,j=1}^{N}b_{ij}\partial_+^2\varphi_i\partial_+\varphi_j
+\sum_{i=1}^{N}c_{i}\partial_+^3\varphi_i.\label{W3}
\eeq
Differentiating (\ref{W3}) by $\partial_-$ and using the equation of motion
(\ref{EofM}), the current-conservation equation
$\partial_-W^{(3)}=0$
is reduced to
\beqa
&&b_{ji}+\beta c_iK_{ij}=0,\n
&&(3a_{ijk}+\beta b_{ij}K_{ik}+\beta^2c_iK_{ij}K_{ik})
+(j\leftrightarrow k)=0,\n
&&a_{ijk}=a_{jik}\label{cceq}
\eeqa
for any $i,j,k$. There are two possible cases:\\
(i)~~$a_{ijk}$, $b_{ij}$ ($i,j,k=1,2$) are written in terms of two
independent coefficients $c_1$ and $c_2$.\\
(ii)~~$a_{ijk}$, $b_{ij}$ ($i,j,k=1,2$) and $c_1$ are written
in terms of $c_2$ only.\\
Note that the derivative of the energy-momentum tensor
$\partial_+T$ is always conserved, its coefficients
satisfying (\ref{cceq}). Since we are looking for spin-3 conserved
currents other than $\partial_+T$, we may set one of $c_i$
to 0 from the outset. Therefore in the case (ii) all $a_{ijk}$,
$b_{ij}$ and $c_i$ are zero, i.e. there are no other conserved
currents, while in the case (i) the existence of another non-vanishing
spin-3 conserved current is indicated.

For rank-2 Kac-Moody algebras with the Cartan matrix (\ref{rank2}),
the solution of (\ref{cceq}) is given by
\beq
c_1=\frac{l-kl}{k-k^2}c_2~~~(k\neq 1,~ k>0)
{}~~~\mbox{
or}~~~
c_1=\frac{l^2-l}{1-l}c_2~~~(k=1).
\eeq
If $k\neq 1$ or $k=1,~l\neq 1$, $c_i$ are related linearly, and
hence there are no conserved currents. The case $k=l=1$
corresponds to $A_2$ (su(3)).

\subsection{Spin 4}

We assume the form of the current as
\beqa
W^{(4)}&\equiv&
\sum_{i,j,k,l=1}^{N}
a_{ijkl}\partial_+\varphi_i\partial_+\varphi_j\partial_+\varphi_k
\partial_+\varphi_l
+\sum_{i,j,k=1}^{N}b_{ij}\partial_+^2\varphi_i\partial_+\varphi_j
\partial_+\varphi_k\n
&&
+\sum_{i,j=1}^{N}(c_{ij}\partial_+^2\varphi_i\partial_+^2\varphi_j
+d_{ij}\partial_+^3\varphi_i\partial_+\varphi_j)
+\sum_{i=1}^{N}e_{i}\partial_+^4\varphi_i.\label{T4}
\eeqa
$\partial_-W^{(4)}=0$ is equivalent to
\beqa
&&d_{ji}+\beta e_iK_{ij}=0,\n
&&2b_{jik} +2\beta c_{ij}K_{ik} +\beta d_{ik}K_{ij}
+3\beta^2 e_i K_{ij}K_{ik}=0,\n
&&(4a_{ijkl}+\beta b_{ijk}K_{il}+\beta^2 d_{ij}K_{ik}K_{il})
+(j\leftrightarrow l)+(j\leftrightarrow k)=0,\n
&&a_{ijkl}=a_{jikl},\n
&&b_{jik}=b_{kij}
\label{W4cceq}
\eeqa
for any $i,j,k,l$.

\hskip -2mm
For TFTs based on rank-2 Kac-Moody algebras
a calculation using MATHEMATICA shows that if
\beq
k>0,~~l>0,~~A\equiv 16-10k-10l-2kl-3k^2l-3kl^2\neq 0,
\eeq
the equations (\ref{W4cceq}) are solved as
\beqa
c_{11}&=&
c_{22}\frac{l^2}{k^2}+e_2\left(
\frac{4l-3lk-l^2+2l^2k}{k^2}\right.\n
&&\left.
+\frac{
\mbox{\scriptsize $
-10l-\frac{64l}{k^2}+\frac{56l}{k}+6l^2
+\frac{88l^2}{k^2}-\frac{64l^2}{k}+3kl^2+16l^3
-\frac{30l^3}{k^2}+\frac{2l^3}{k}-6kl^3-6l^4+\frac{9l^4}{k}$
}
}
{
A}
\right)\n
c_{12}&=&
c_{22}\frac{l}{k}+e_2\left(
\frac{7l-3kl}{4}+\frac{3l}{2k}\right.\n
&&\left.
-\frac{\mbox{\scriptsize
$17l+\frac{8l}{k}+\frac{39kl}{2}+\frac{15k^2l}{2}+\frac{13l^2}{2}
-\frac{5l^2}{k}-\frac{9kl^2}{2}-\frac{15k^2l^2}{4}
+\frac{9k^3l^2}{4}+\frac{3l^3}{2}-\frac{9kl^3}{4}+\frac{9k^2l^3}{4}$}
}{
A}
\right),\n
e_1&=&e_2\left(
\frac{
-10l+\frac{16l}{k}-2l^2-\frac{10l^2}{k}+3kl^2+3l^3
}
{
A}
\right).\label{Mathsol}
\eeqa
Hence the solution is expressed by two parameters $c_{22}$ and $e_2$.
Since $W^{(4)}$ includes two trivial conserved currents
$\partial_+^2 T$ and $(T)^2$, two parameters can be set to zero
from the beginning. This means that (\ref{Mathsol}) corresponds to the
cases with no conserved currents. Hence the necessary condition for
such a current to exist is
$A=0$.
This together with the condition (\ref{kl>4}) shows
\beq
-4<k+l<\frac{2}{3}. \label{k+l}
\eeq
It is clear that there are no $k,l$ which satisfy both (\ref{kl>4})
and (\ref{k+l}). This establishes the non-existence of non-trivial
spin-4 conserved currents for HTFTs
based on rank-2 Kac-Moody algebras.

For SHTFTs based on rank-3 and -4 Kac-Moody algebras we can also
explicitly check that the solution of (\ref{W4cceq}) is parameterized
by only two parameters, and hence they have no conserved currents,
either.

\section{Summary and prospects}

We have shown that the Painlev\'{e} test is useful not only for
probing (non-) integrability but also for finding the values of
spins of conserved currents in TFTs. The locations
of resonances precisely give the spins of W currents for TFTs
based on simple Lie algebras.
We applied this test to SHTFTs, and showed that there exists no
resonance except for the one at $n=2$, which corresponds to the
energy-momentum tensor,
indicating the non-integrability. As a check, we have
explicitly seen that there are no spin-3 nor -4 conserved currents
for these theories.

One might think that the conformal invariance in two dimensions and
the integrability are not compatible, since the former may lead to an
infinite number of conserved charges. Our interpretation of this
`discrepancy' is as follows: Despite being infinite, the number of conserved
charges for (S)HTFTs may not be sufficient to be integrable. Namely, an
(S)HTFT contains as many fields as the number of rank $N(\geq 2)$, hence
there are `$N\times\infty$' independent modes, while a single
conserved Virasoro current provides only `$1\times\infty$' charges.
We also recall an analogy in the relation between conformal field
theories and integrable lattice models; the minimal series
(i.e. $c<1$) of the former correspond to the latter on criticality, but
there is no such correspondence if $c(=\mbox{the number of degrees of
freedom})$ is larger than 1. In view of this, not all conformal field
theory in two dimensions may necessarily be integrable if the number
of fields$>1$, unless other additional symmetries (e.g.
W currents, Kac-Moody currents) exist in the system.

Finally, we will comment on general HTFTs.
Although (\ref{na=nb+1=2}) is not the unique situation
for general cases, there is no difficulty to perform the test for
these theories in practice, starting from (\ref{allAnonzero}).
We have also checked for all rank-3, -4 and -5 HTFTs that the resonance
always occurs at $n=2$ only. In view of this fact and our
conserved-current analysis, we conjecture that all the HTFTs are
non-integrable. One of the hints to clarify this point will be Ziglin's
theorem, which was already used to show the integrability of some
systems by Yoshida et. al. \cite{Yoshida_Ziglin}. The cohomological
analysis of Feigin and Frenkel may give us another suggestion on this
problem \cite{FF}. It will also be interesting
to establish the relation, if any, between the formal non-integer
resonances and higher order Casimirs for generalized Kac-Moody algebras
\cite{KacCasimir}.

\appendix
\section*{Appendix A}

In this appendix we prove that (\ref{na=nb+1=2}) is the only
possibility for TFTs based on simple Lie algebras or strictly
hyperbolic Kac-Moody algebras.
Let us assume $n_B\geq 2$.
Substituting the expansion (\ref{ABexpansion}) in (\ref{ls}),
we have
\beq
\sum_{m=0}^{n+n_B-1}A_i^{(n-m+n_B-1)}\sum_{j=1}^N K_{ij}B_j^{(m)}
=\left\{\begin{array}{l}
(n-n_A)\Dot{\phi}A_i^{(n)}+\Dot{A}_i^{\raisebox{-1.5mm}{\scriptsize $(n-1)$}}
{}~~~\mbox{if $n\geq 0$},\\
0
{}~~~\mbox{if $-n_B+1\leq n\leq -1$}
\end{array}
\right.\label{B2}
\eeq
and
\beq
\sum_{n=0}^{\infty}\left\{
(n-n_B)\phi'B_j^{(n)}+{B_j^{(n-1)}}'
\right\}\phi^{n-n_B-1}
=
-\sum_{n=0}^{\infty}A_j^{(n)}\phi^{n-n_A}, \label{B3}
\eeq
where $A_j^{(-1)}\equiv B_j^{(-1)}\equiv 0$. We may, without loss of
generality, assume that not all $A_i^{(0)}$ and not all $B_j^{(0)}$
are zero (because if so, we may then redefine $n_A\rightarrow n_A+1$,
etc.). (\ref{B3}) then means
\beq
n_B+1=n_A,\label{B4}
\eeq
and hence
\beq
(n-n_B)\phi'B_j^{(n)}+{B_j^{(n-1)}}'=-A_j^{(n)}.
\label{B5}
\eeq
On the other hand, we find from (\ref{B2}) that
\beq
A_i^{(0)}\sum_{j=1}^N K_{ij}B_j^{(0)}=0~~~(i=1,\ldots,N).
\label{B6}
\eeq
If
\beq
\mbox{$A_i^{(0)}\neq 0$ for any $i=1,\ldots,N $}
{}~~~\mbox{(lowest balance)},
\eeq
then
\beq
B_1^{(0)}=B_2^{(0)}=\cdots=B_N^{(0)}=0,
\eeq
which contradicts the assumption ($K$ is assumed to be invertible.).
So let
\beq\
A_1^{(0)}\neq 0,\ldots, A_P^{(0)}\neq 0,~~
A_{P+1}^{(0)}=\cdots =A_N^{(0)}=0.\label{B9}
\eeq
Substituting (\ref{B9}) into (\ref{B5}), we have
\beq
B_1^{(0)}\neq 0,\ldots, B_P^{(0)}\neq 0,~~
B_{P+1}^{(0)}=\cdots =B_N^{(0)}=0.\label{B10}
\eeq
Substituting (\ref{B9})(\ref{B10}) into (\ref{B6}), we obtain
\beq
\sum_{j=1}^P K_{ij}B_j^{(0)} =0~~~(i=1,\ldots,P).
\label{B11}
\eeq
If the $P\times P$ minor $\{K_{ij; ~ i,j-1,\ldots,P}\}$ is invertible,
then $B_1^{(0)}=\cdots =B_P^{(0)}=0$, which contradicts (\ref{B10}).
Hence it may not have its inverse. This cannot be satisfied by $K_{ij}$
corresponding to simple Lie algebras or strictly hyperbolic Kac-Moody
algebras, and the proof is thus completed.

\section*{Appendix B}
In this Appendix we give a proof of the Theorem in sect.3.
Here we show explicitly that the eigenvalues indeed coincide
to the set of numbers known as the exponents, case by case. For $A_N,
B_N, C_N, D_N$ series we employ induction.
\begin{itemize}
\item{$A_N$.}
\end{itemize}

The Cartan matrix and the associated $D$ matrix are given by
\beqa
K^{(N)}&=&
\left[
\begin{array}{crclc}
2&-1&&&\\
{}~-1&2&~-1&&\\
&\ddd&\ddd&\ddd&\\
&&-1~&2&-1~{}\\
&&&-1&2
\end{array}
\right],\\
D^{(N)}&=&\mbox{Diag}\left[
\frac{N\cdot 1}{2},\frac{(N-1)\cdot 2}{2}, \ldots,
\frac{(N-i+1)\cdot i}{2},\ldots,
\frac{1\cdot N}{2}
\right].
\eeqa

Induction. If $N=1$, then $2K^{(1)}D^{(1)}=2$, which trivially satisfies
the Theorem.
We next assume that $2K^{(N-1)}D^{(N-1)}$ has the spectrum
$\{2,6,10, \ldots,\\
N (N-1) \}$. The following lemma shows that the matrix
$2K^{(N)}D^{(N)}$ also has the same eigenvalues:
\newtheorem{lemmaA}{Lemma A}{\bf}{\it}
\begin{lemmaA}
The matrix
\beq
P^{(N)}\equiv\underbrace{
\left[
\begin{array}{ccccc}
N-1&&&&\\
1&N-2&&&\\
&2&\ddd&&\\
&&\ddd&2&\\
&&&N-2&1\\
&&&&N-1
\end{array}
\right]
}_{N-1}\left.\rule{0mm}{15mm}\right\}\mbox{\scriptsize $N$}
\eeq
is an `intertwiner' of $2K^{(N)}D^{(N)}$ and $2K^{(N-1)}D^{(N-1)}$,
i.e. it satisfies
\beq
2K^{(N)}D^{(N)}\cdot P^{(N)}=P^{(N)}\cdot 2K^{(N-1)}D^{(N-1)}.
\eeq
\end{lemmaA}
\vskip 5mm\noindent
{\it Proof.}~~~
Easy.
\vskip 5mm

Therefore, to complete induction, we have only to prove that
the extra eigenvalue of $2K^{(N)}D^{(N)}$ is $N(N+1)$. This can be
shown by the following Lemma:
\begin{lemmaA}
\beq
v=\left[
1,-\frac{N}{2}, \frac{N(N-1)}{6},\ldots,
(-1)^{i+1}\frac{N!}{i!(N-i+1)!},
\ldots,
(-1)^N\frac{N}{2}, (-1)^{N+1}
\right]^T
\eeq
is an eigenvector of $2K^{(N)}D^{(N)}$ with the eigenvalue $N(N+1)$.
\end{lemmaA}
\vskip 5mm\noindent
{\it Proof.}~~~
Straightforward.
\vskip 5mm

This completes the proof of the Theorem for the $A_N$ type.

\begin{itemize}
\item{$B_N$, $C_N$.}
\end{itemize}

For these types one can prove the Theorem in the same way
as we have done for $A_N$.

For $B_N$ the Cartan matrix and the $D$ matrix are
\beqa
K^{(N)}&=&
\left[
\begin{array}{crclcc}
2&-1&&&&\\
{}~-1&2&~-1&&&\\
&\ddd&\ddd&\ddd&&\\
&&-1~&2&-1~{}&\\
&&&-1&2&-2\\
&&&&-1&2
\end{array}
\right],\\
(D^{(N)})_{ij}&=&\left\{
\begin{array}{l}
i\left(N-\frac{i}{2}+\frac{1}{2}\right)\delta_{ij}~~~(1\leq i\leq N-1),\\
\frac{1}{4}N(N+1)\delta_{Nj}~~~(i=N).
\end{array}
\right.
\eeqa

The assertion can be easily checked for $N=2$. Making use of the
following two Lemmas enables us to show, similarly by induction,
that the matrix $2K^{(N)}D^{(N)}$ has the eigenvalues
$\{1\cdot 2,~ 3\cdot 4,~ 5\cdot 6,\ldots,~ (2N-1)\cdot 2N \}$.
\newtheorem{lemmaB}{Lemma B}{\bf}{\it}
\begin{lemmaB}
The $N\times (N-1)$ matrix $P^{(N)}$, given by
\beq
(P^{(N)})_{ij}=\left\{
\begin{array}{l}
(-1)^{i+j}\frac{\raisebox{1mm}{$2N (2N-2j-1) j! (2N-j-1)!$}}
{\raisebox{-1mm}{$i! (2N-i+1)!$}}\\
{}~~~~~~~~~~~~(1\leq i\leq j\leq N-1),\\
\\
\frac{\raisebox{1mm}{$j$}}{\raisebox{-1mm}{$2N-j$}}~~~(2\leq i =j+1\leq N),\\
\\
0~~~\mbox{(otherwise)},
\end{array}
\right.
\eeq
satisfies
\beq
2K^{(N)}D^{(N)}\cdot P^{(N)}=P^{(N)}\cdot 2K^{(N-1)}D^{(N-1)}.
\eeq
\end{lemmaB}
\begin{lemmaB}
A column vector
\beq
v^{(N)}_i=\frac{(-1)^{i}(2N-2i+1)}{i!(2N-i+1)!}~~~(i=1,\ldots, N)
\eeq
is an eigenvector of $2K^{(N)}D^{(N)}$ of the eigenvalue $(2N-1)2N$.
\end{lemmaB}

Both Lemmas can be verified by straightforward calculations.

For $C_N$, the necessary informations are as follows:
\beqa
K^{(N)}&=&
\left[
\begin{array}{crclcc}
2&-1&&&&\\
{}~-1&2&~-1&&&\\
&\ddd&\ddd&\ddd&&\\
&&-1~&2&-1~{}&\\
&&&-1&2&-1\\
&&&&-2&2
\end{array}
\right],\\
(D^{(N)})_{ij}&=&\left\{
\begin{array}{l}
i\left(N-\frac{i}{2}\right)\delta_{ij}~~~(1\leq i\leq N-1),\\
\frac{1}{2}N^2\delta_{Nj}~~~(i=N).
\end{array}
\right.
\eeqa
\newtheorem{lemmaC}{Lemma C}{\bf}{\it}
\begin{lemmaC}
\beq
2K^{(N)}D^{(N)}\cdot P^{(N)}=P^{(N)}\cdot 2K^{(N-1)}D^{(N-1)},
\eeq
where
\beq
(P^{(N)})_{ij}=\left\{
\begin{array}{l}
(-1)^{i+j}\frac{\raisebox{1mm}{$2N(2N-1)(N-i)\cdot j!(2N-2-j)!          $}}
{\raisebox{-1mm}{$(N-1)\cdot i!(2N-i)!              $}}\\
{}~~~~~~~~~~~~~~~~~~~~~~~~~~~~(1\leq i\leq j\leq N-2),\\
\\
\frac{\raisebox{1mm}{$Nj$}}
{\raisebox{-1mm}{$(N-1)(2N-j-1)$}}~~~
(2\leq i =j+1\leq N),\\
\\
(-1)^{N+i-1}\frac{\raisebox{1mm}{$(2N-1)(N-i)\cdot N!(N-2)!          $}}
{\raisebox{-1mm}{$i!(2N-i)! $}}\\
{}~~~~~~~~~~~~~~~~~~~~~~~~~~~~(j=N-1, 1\leq i\leq N-1),\\
\\
0~~~\mbox{(otherwise)}.
\end{array}
\right.
\eeq
\end{lemmaC}
\begin{lemmaC}
A column vector
\beq
v^{(N)}_i=\frac{(-1)^{i}}{i!(2N-i)!}~~~(i=1,\ldots, N)
\eeq
is an eigenvector of $2K^{(N)}D^{(N)}$ of the eigenvalue $(2N-1)2N$.
\end{lemmaC}

The proof is completely parallel,
and we leave it to the reader.

\begin{itemize}
\item{$D_N$.}
\end{itemize}

The Cartan matrix and the $D$ matrix of the $D_N$ type Lie
algebra are
\beqa
K^{(N)}&=&
\left[
\begin{array}{crclccc}
2&-1&&&&&\\
{}~-1&2&~-1&&&&\\
&\ddd&\ddd&\ddd&&&\\
&&-1~&2&-1~{}&&\\
&&&-1&2&-1&-1\\
&&&&-1&2&0\\
&&&&-1&0&2
\end{array}
\right],\\
(D^{(N)})_{ij}&=&\left\{
\begin{array}{l}
i\left(N-\frac{i}{2}-\frac{1}{2}\right)\delta_{ij}~~~(1\leq i\leq N-2),\\
\frac{1}{4}N(N-1)\delta_{ij}~~~(i=N-1, N).
\end{array}
\right.
\eeqa
What we have to show is that the matrix $2 X^{(N)}\equiv
2K^{(N)}D^{(N)}$ has eigenvalues
$\{1\cdot2,~3\cdot 4,\ldots,~ (2N-3)(2N-2);~ (N-1)N\}$ for any
$N=4,5,\ldots$. Due to the `middle' eigenvalue $(N-1)N$, we need some
preparation before applying induction to this case.

Let us consider a symmetric $Y^{(N)}\equiv (D^{(N)})^{\frac{1}{2}}
K^{(N)}(D^{(N)})^{\frac{1}{2}}$, which has the same set of eigenvalues as
$X^{(N)}$ (Here we have already used the assumption of induction in
anticipating reality of the square root of $D^{(N)}$).
It is easy to see that
\newtheorem{lemmaD}{Lemma D}{\bf}{\it}
\begin{lemmaD}
$u^{(N)}\equiv [0,\ldots,0,-1,1]^T$ is an eigenvector of
$2Y^{(N)}$ of the eigenvalue $N(N-1)$.
\end{lemmaD}

This shows that $2X^{(N)}$ also has the eigenvalue $N(N-1)$.
Since $Y^{(N)}$ is symmetric,
\beq
\frac{1}{2} u^{(N)}(u^{(N)})^T
\eeq
is a projection operator to the vector space spanned by $u^{(N)}$.
Hence, due to the assumption, the matrices
\beq
2{Y^{(N-1)}}'\equiv 2Y^{(N-1)}-\frac{(N-1)(N-2)}{2}u^{(N-1)}(u^{(N-1)})^T
\eeq
and
\beqa
2{X^{(N-1)}}'&\equiv& 2X^{(N-1)}
-(D^{(N)})^{-\frac{1}{2}}
\frac{(N-1)(N-2)}{2}u^{(N-1)}(u^{(N-1)})^T
(D^{(N)})^{\frac{1}{2}}\n
&=&
2X^{(N-1)}
-\frac{(N-1)(N-2)}{2}u^{(N-1)}(u^{(N-1)})^T
\eeqa
have eigenvalues $\{0,~1\cdot 2,\ldots,~(2N-5)(2N-4)\}$.

It now suffices to show that $2{X^{(N)}}'$ also possesses the same
spectrum except one extra eigenvalue $(2N-3)(2N-2)$, as well as
that, for the $D_4$ case, $2{X^{(4)}}'$ has eigenvalues
$\{0, 2, 12, 30\}$. The latter can be done easily. To show the former,
we can take the same steps as the previous proofs for other types
of Lie algebras.

\begin{lemmaD}
The $N\times (N-1)$ matrix $P^{(N)}$, given by
\beq
(P^{(N)})_{ij}=\left\{
\begin{array}{l}
(-1)^{i+j}\frac{\raisebox{1mm}{$2N (N-1) (2N-2j-3) j! (2N-j-3)!$}}
{\raisebox{-1mm}{$(N-2) i! (2N-i-1)!$}}\\
{}~~~~~~~~~~~~~~~~~~~~~~~~~~~~(1\leq i\leq j\leq N-3),\\
\\
(-1)^{N+i}\frac{\raisebox{1mm}{$N((N-1)!)^2$}}
{\raisebox{-1mm}{$(N-2)i!(2N-i-1)!$}}\\
{}~~~~~~~~~~~~~~~~~~~~~~~~~~~~(j=N-2, N-1;~~1\leq i\leq N-2 ),\\
\\
\frac{\raisebox{1mm}{$jN$}}
{\raisebox{-1mm}{$(N-2)(2N-j-2)$}}~~~
(2\leq i =j+1\leq N-2),\\
\\
\delta_{ij}~~~(i=N-1,N;~~j=N-2,N-1),
\\
0~~~\mbox{(otherwise)}
\end{array}
\right.
\eeq
satisfies
\beq
2{X^{(N)}}'\cdot P^{(N)}=P^{(N)}\cdot 2{X^{(N-1)}}'.
\eeq
\end{lemmaD}

One can prove the above by a straightforward calculation in the same
manner.

The following Lemma completes the proof of the $D_N$ case:

\begin{lemmaD}
$2X^{(N)}$ has an eigenvalue $(2N-3)(2N-2)$.
\end{lemmaD}
\vskip 5mm\noindent
{\it Proof.}~~~
It is easy to verify that the column vector
\beq
v^{(N)}_i\equiv\left\{
\begin{array}{l}
\frac{\raisebox{1mm}{$(-1)^i(2N-2i-1)$}}
{\raisebox{-1mm}{$i!(2N-i-1)!$}}~~~(1\leq i\leq N-2),\\
\\
\frac{\raisebox{1mm}{$(-1)^{N+1}$}}
{\raisebox{-1mm}{$(N-1)!N!$}}~~~(i= N-1,N)
\end{array}
\right.
\eeq
satisfies
\beq
2X^{(N)}v^{(N)}=(2N-3)(2N-2)v^{(N)}.
\eeq q.e.d.
\vskip 5mm
\begin{itemize}
\item{Exceptional types.}
For the exceptional types of simple Lie algebras one can
explicitly establish the validity of the Theorem. We will
list the eigenvectors for these cases for completeness:

\begin{itemize}
\item{$E_6$.}
\end{itemize}
\beq
K=\left[\begin{array}{cccccc}
2&-1&0&0&0&0\\
-1&2&-1&0&0&0\\
0&-1&2&-1&-1&0\\
0&0&-1&2&0&0\\
0&0&-1&0&2&-1\\
0&0&0&0&-1&2
\end{array}\right],\eeq
\beq
D=\mbox{Diag}\left[
8, 15, 21, 11, 15, 8
\right],
\eeq
\beq
P=
\left[
\begin{array}{cccccc}
1& 1& 1& 1& 1& 1 \\
-1& -2/5& 0& 0&2/5& 1\\
1& 1/15& -1/3& -1& 1/15& 1\\
1& -4/5& -16/35& 8/5& -4/5& 1\\
-1& 4/3& 0& 0& -4/3& 1\\
1& -10/3& 16/3& -28/11& -10/3& 1
\end{array}
\right]^T,
\eeq
\beq
P^{-1}2KDP=\mbox{Diag}\left[
2,20,30,56,72,132
\right].
\eeq

\begin{itemize}
\item{$E_7$.}
\end{itemize}

\beq
K=\left[\begin{array}{ccccccc}
2&-1&0&0&0&0&0\\
-1&2&-1&0&0&0&0\\
0&-1&2&-1&0&0&0\\
0&0&-1&2&-1&-1&0\\
0&0&0&-1&2&0&0\\
0&0&0&-1&0&2&-1\\
0&0&0&0&0&-1&2
\end{array}\right],
\eeq
\beq
D=
\left[
\frac{27}{2}, 26, \frac{75}{2}, 48, \frac{49}{2}, 33, 17
\right],
\eeq
\beq
P=
\left[
\begin{array}{ccccccc}
1& 1& 1& 1& 1& 1& 1\\
-65/33& -10/11& -31/165&17/66& 4/11&19/33&1\\
13/22 &-1/44 & -5/22 &-37/176&-37/77 &2/11&1\\
-65/9&5&53/15&-1/2&-6& -1/3&1\\
-5/12&5/8&-1/12&-17/48&1&-32/33&1\\
10/11&-320/143&2&6/11&-48/77&-19/11&1\\
-34/99&238/143&-170/39&68/11&-408/143&-119/33&1
\end{array}
\right]^T,
\eeq
\beq
P^{-1}2KDP=\mbox{Diag}\left[
2,30,56,90,132,182,306
\right].
\eeq

\begin{itemize}
\item{$E_8$.}
\end{itemize}

\beq
K=\left[\begin{array}{cccccccc}
2&-1&0&0&0&0&0&0\\
-1&2&-1&0&0&0&0&0\\
0&-1&2&-1&0&0&0&0\\
0&0&-1&2&-1&0&0&0\\
0&0&0&-1&2&-1&-1&0\\
0&0&0&0&-1&2&0&0\\
0&0&0&0&-1&0&2&-1\\
0&0&0&0&0&0&-1&2
\end{array}
\right],
\eeq
\beq
D=\mbox{Diag}\left[
29, 57, 84, 110,  135, 68,  91, 46
\right],
\eeq
\beq
P
=\left[\mbox{
\scriptsize$
\begin{array}{cccccccc}
1& 1& 1& 1& 1& 1& 1& 1\\
-38/13&-20/13&-103/182&6/13&15/13& 15/26& 64/91& 1\\
228/245& -32/245&  -97/245&  -807/2695&  -2/21&  -9/49& 2 /7& 1\\
-19/15& 11/15& 67/105& 1/15& -1/3& -1& 1/91& 1\\
-33/13& 55/13& -99/91& -213/91& -397/819& 5955/1547& - 61/91& 1\\
4/13& -176/247& 7/13& 17/65& -18/65& 9/13& -14/13& 1\\
-874/1911&10028/5733&-5267/1638&4301/1911&874/819&-1311/1274&-184/91&1\\
437/4901& -23/39& 1127/5507& -4301/845& 437/65& -513/169& -49/13& 1
\end{array}$}
\right]^T,
\eeq
\beq
P^{-1}2KDP=\mbox{Diag}\left[
2,56,132,182,306,380,552,870
\right].
\eeq

\begin{itemize}
\item{$F_4$.}
\end{itemize}

\beq
K=\left[\begin{array}{cccc}
2&-1&0&0\\
-1&2&-2&0\\
0&-1&2&-1\\
0&0&-1&2
\end{array}
\right],
\eeq
\beq
D=\mbox{Diag}\left[11,  21,  15,  8\right],
\eeq
\beq
P=
\left[
\begin{array}{cccc}
1& 1& 1& 1\\
-1& -1/3& 1/15& 1\\
8/5& -16/35& -4/5& 1\\
-28/11& 16/3& -10/3& 1
\end{array}
\right]^T,
\eeq
\beq
P^{-1}2KDP=\mbox{Diag}\left[
2,30,56,132
\right].
\eeq

\begin{itemize}
\item{$G_2$.}
\end{itemize}

\beq
K=\left[\begin{array}{cc}
2&-3\\
-1&2
\end{array}\right],
\eeq
\beq
D=\mbox{Diag}\left[
5,3
\right],
\eeq
\beq
P=
\left[
\begin{array}{cc}
1& 1\\-9/5& 1
\end{array}
\right]^T,
\eeq
\beq
P^{-1}2KDP=\mbox{Diag}\left[
2,30
\right].
\eeq
\end{itemize}

We have thus proven the Theorem for all the types of simple Lie algebras.

\subsection*{Acknowledgment}
We would like to thank H. Yoshida for critical remarks and useful
suggestions, which helped us complete the final version of this article.
We are grateful to R. Sasaki for a careful reading of the manuscript
and to Y. Yamada for illuminating discussions. We also thank
M. Jimbo, H. Nicolai and K. Takasaki for discussions.

\newpage
\begin{table}
  \caption{Exponents for simple Lie algebras.}
 \centering
     \begin{tabular}{cc}
        \hline\noalign{\smallskip}
    \mbox{\boldmath $g$} & Exponents
\\
        \noalign{\smallskip}
        \hline
        \noalign{\smallskip}
&\\
$A_N$ &$ 1,2,3,\ldots,N$\\
$B_N$ $(C_N)$ &$ 1,3,5,\ldots,2N-1$\\
$D_N$ & $1,3,5\ldots,2N-3~\mbox{and}~N-1$\\
$E_6$&$1,4,5,7,8,11$\\
$E_7$&$1,5,7,9,11,13,17$\\
$E_8$&$1,7,11,13,17,19,23,29$\\
$F_4$&$1,5,7,11$\\
$G_2$&$1,5$\\\noalign{\smallskip}
               \hline
               \noalign{\smallskip}
     \end{tabular}
\end{table}

\newpage
\setcounter{table}{2}

\begin{table}
  \caption{Eigenvalues of $2KD$ for strictly
hyperbolic Kac-Moody algebras}
 \centering
     \begin{tabular}{ccccc}
        \hline\noalign{\smallskip}
    Rank & Number& Eigenvalues
& Number& Eigenvalues
\\
        \noalign{\smallskip}
        \hline
        \noalign{\smallskip}
&&&&\\
        3
& 1 &$2,-12,-12$
& 1d &$2,-9,-15$
\\&&&&\\
        \noalign{\smallskip}
& 2 & $2,-6,-24$
& 2d &$2,-10,-20$
\\&&&&\\
& 3 &$2,-\frac{15}{2},-\frac{15}{2}$
& 3d &$2,-\frac{9}{2},-\frac{21}{2}$
\\&&&&\\
        \noalign{\smallskip}
& 4 &$2,-12,-42$
& 4d &$2,\frac{-54 \pm 2 \sqrt{113}}{2}$
\\&&&&\\
        \noalign{\smallskip}
& 5 &$2,\frac{-58\pm 2\sqrt{281}}{2}$
& 5d &$2,\frac{-58\pm 2\sqrt{181}}{2}$
\\&&&&\\
        4
& 1 &$2,-12,\frac{-30\pm 2\sqrt{57}}{2}$
&&\\&&&&\\               \noalign{\smallskip}
               \hline
               \noalign{\smallskip}
     \end{tabular}
\end{table}
\end{document}